\documentclass[aps,prl,showpacs,superscriptaddress,twocolumn]{revtex4}
\usepackage{amsfonts}
\usepackage{subfigure}
\usepackage{txfonts}
\usepackage{amssymb}
\usepackage{amsbsy} 
\usepackage{epsfig}

\def\be{\begin{equation}} \def\ee{\end{equation}}
\def\bea{\begin{eqnarray}} \def\eea{\end{eqnarray}}

\def\bq{{\bf q}}

\def\bk{{\bf k}}

\def\bK{{\bf K}}
\def\be{{\bf e}}
\def\bd{{\bf d}}
\def\bA{{\bf A}}

\begin{document}

\title{ Tunable Weyl Points in Periodically Driven Nodal Line Semimetals  }

\author{Zhongbo Yan}
\altaffiliation{ yzhbo@mail.tsinghua.edu.cn} \affiliation{ Institute for
Advanced Study, Tsinghua University, Beijing, 100084, China}

\author{Zhong Wang}
\altaffiliation{ wangzhongemail@tsinghua.edu.cn} \affiliation{ Institute for
Advanced Study, Tsinghua University, Beijing, 100084, China}

\affiliation{Collaborative Innovation Center of Quantum Matter, Beijing 100871, China }


\begin{abstract}

Weyl semimetals and nodal line semimetals are characterized by linear
band touching at zero-dimensional points and one-dimensional lines,
respectively. We predict that a circularly polarized light drives
nodal line semimetals into Weyl semimetals. The Floquet Weyl points
thus obtained are tunable by the incident light, which enables
investigations of them in a highly controllable manner. The
transition from nodal line semimetals to Weyl semimetals is
accompanied by the emergence of a large and tunable anomalous Hall
conductivity. Our predictions are experimentally testable by
transport measurement in film samples or by pump-probe angle-resolved
photoemission spectroscopy.

\end{abstract}


\maketitle

It has become well known that topological concepts underlie many fascinating phenomena in
condensed matter physics. After in-depth investigations of
topological insulators\cite{hasan2010,qi2011,Chiu2015RMP},
considerable attention is now focused on topological semimetals.
Unlike topological insulators, whose gapless excitations always live
at the sample boundary, topological semimetals host gapless fermions in
the bulk. The two major classes of topological semimetals under
intense study are (i) nodal point semimetals and (ii) nodal line
semimetals (NLSM). The nodal point semimetals include Dirac
semimetals(DSM)\cite{liu2014discovery,neupane2014,Borisenko2014,xu2015observation,
wang2012dirac,young2012dirac,wang2013three,Sekine2014,Zhang2015detection,
Chen2015Magnetoinfrared,Yuan2015ZrTe5} and Weyl
semimetals(WSM)\cite{wan2011,nielsen1983adler,volovik2003,yang2011,
burkov2011,weng2015,Huang2015TaAs,Zhang2015a,Xu2015weyl,lv2015,Huang2015,
YangLexian,Ghimire,Shekhar,Xu2015NbAs,lu2015weyl,lu2013weyl}. The main
feature of the band structures of DSMs and WSMs is the linear
band-touching points (``Dirac points'' and ``Weyl points''), which
are responsible for most of their interesting properties, including
novel phenomena induced by the chiral
anomaly\cite{son2012,liu2012,aji2011,zyuzin2012,wang2013a,Hosur-anomaly,
Hosur2013,Kim-chiral-anomaly,Parameswaran-anomaly,
Zhou-plasmon,Li2015ZrTe5,Bi2015,Goswami2015}.
NLSMs\cite{Burkov2011nodal,Carter2012,
Phillips2014tunable,chen2015topological,
Zeng2015nodal,Chiu2014,Mullen2015,Weng2015nodal,Yu2015,Kim2015,Bian2015nodal,
Xie2015ring,Rhim2015Landau,Chen2015spin,
Fang2015nodal,Bian2015TlTaSe,Chan2015,Rhim2016,Yan2016nodal} differ in that they contain
band-touching \emph{lines} or \emph{rings}\footnote{ For a simple
narrative, we do not distinguish between the terms ``nodal line''
and ``nodal ring'' hereafter.}, away from which the dispersion is
linear.

In this Letter we show that driving NLSMs by a circularly polarized
light (CPL) creates WSMs, namely, nodal lines become nodal points
under radiations. Our work was motivated by recent progress in
Floquet topological states
\cite{lindner2011floquet,Kitagawa2011,Oka2009,Inoue2010,
Gu2011,Kitagawa2010a,Kitagawa2010b,
Lindner2013,Jiang2011,rudner2013anomalous,Dahlhaus2011,Gomez2013,
Zhou2011Optical,Delplace2013,wang2013observation,
mahmood2016selective,wang2014floquet,Seetharam2015, Hubener2016,Wang2016Network}, in
particular, Ref.\cite{Chan2016hall} showed that incident light can
shift the locations of Weyl points in WSMs. The effect we
predict in NLSMs is more dramatic:  band-touching lines are driven to points; thus, the dimension of the band-touching manifold
is changed. Meanwhile, a large anomalous Hall conductivity tunable by
the incident light emerges. Unlike the photoinduced Hall effect in
WSMs\cite{Chan2016hall}, which is proportional to intensity of
incident light, the Hall conductivity in our systems is large and
quite insensitive to the light intensity at low temperature, though
it depends sensitively on the incident angle of light. The surface
Fermi arcs of the Floquet WSMs have a simple interpretation, namely,
it comes from tilting the drumhead surface dispersion of NLSMs.

The Floquet WSMs derived from NLSMs are highly tunable, in particular, the Weyl points can be freely tuned to any locations on the nodal line. Hopefully this tunability will motivate further investigations of fascinating properties of topological materials.

Recently, there have appeared experimental evidences of nodal lines in PbTaSe$_2$\cite{Bian2015nodal}, ZrSiS\cite{schoop2015dirac,Neupane2016,Singha2016,Wang2016evidence}, ZrSiTe\cite{hu2016ZrSiTe}, and PtSn$_4$\cite{Wu2016PtSn4}, and quite a few theoretical proposals in Cu$_3$PdN\cite{Yu2015,Kim2015}   Ca$_3$P$_2$\cite{Xie2015ring,Chan2015} Hg$_3$As$_2$\cite{Lu2016Node} and three-dimensional graphene networks\cite{Weng2015nodal}. Thus our prediction can be
experimentally tested in the near future.

{\it Drive nodal line semimetals to Weyl semimetals.---} NLSMs with negligible spin-orbit coupling (SOC) can be regarded as two copies of spinless systems; thus, we  first consider spinless models for notational simplicity. Near the nodal line, the physics can be captured by two-band models\cite{Fang2015nodal,Yu2015,Kim2015}. Our starting point is the model Hamiltonian $\hat{H}=\sum_{k}
\hat{\Psi}_{\bk}^{\dag}\mathcal{H}(\bk)\hat{\Psi}_{\bk}$ with
$\hat{\Psi}_{\bk}=(\hat{c}_{\bk, a},\hat{c}_{\bk, b})^{T}$ and
($\hbar=c=k_{B}=1$),
\begin{eqnarray}
\mathcal{H}(\bk)=[m-Bk^{2}]\tau_{x}+vk_{z}\tau_{z} + \epsilon_{0}(\bk)\tau_0,\label{nlsm}
\end{eqnarray}
where $a,b$ refer to the two orbitals involved and $m$, $B$ are positive constants with the dimension of energy and inverse energy,
respectively; $v$ refers to the Fermi velocity along $z$ direction; $k^{2}=k_{x}^{2}+k_{y}^{2}+k_{z}^{2}$,
and $\tau_{x,y,z}$ are Pauli matrices and $\tau_0$ is the identity matrix. Although quite simple, this two-band model well describes several candidates of NLSMs\cite{Yu2015,Kim2015} in which the spin-orbit coupling can be neglected.  The form of $\epsilon_0(\bk)$ is not crucial and is specified later.
When $\epsilon_0=0$, the energy spectra of this Hamiltonian read
\begin{eqnarray}
E_{\pm,\bk}=\pm\sqrt{[m-Bk^{2}]^{2}+v^{2}k_{z}^{2}}.
\end{eqnarray}
The nodal ring, on which the two bands touch, is located at the $k_z=0$ plane and determined by the equation $k_{x}^{2}+k_{y}^{2}=m/B$. The nodal ring
is protected by a mirror symmetry, $\mathcal{M}\mathcal{H}(k_{x,y},k_{z})\mathcal{M}^{-1}
=\mathcal{H}(k_{x,y},-k_{z})$, with $\mathcal{M}=i \tau_{x}$\footnote{In general,  band touching is protected by the combination of topology and symmetry\cite{horava2005,Zhao2013Topological}.}.

\begin{figure}
\includegraphics[width=8.0cm, height=3.5cm]{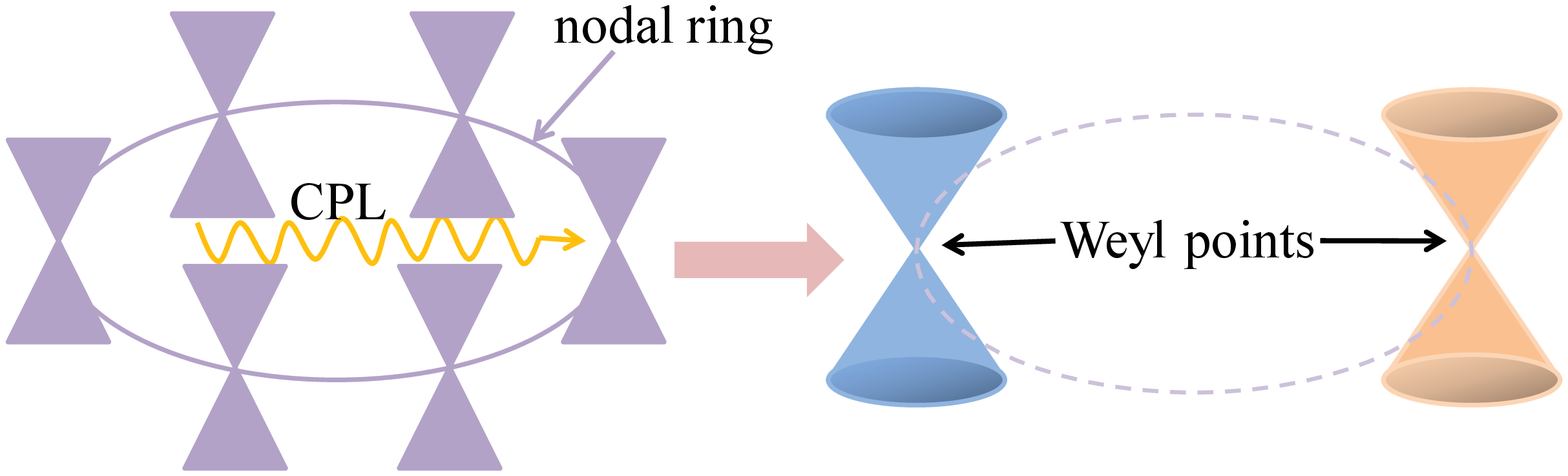}
\caption{ Conceptual illustration. A Dirac nodal line  semimetal has
band-touching lines/rings and linear dispersion in the two transverse
directions.  CPL lifts all the
band touching except at two isolated points, near which the
dispersion of the Floquet bands is linear in all three directions,
i.e., Weyl points. The different colors of the Weyl cones represent
different chirality ($\pm 1$).}  \label{sketch}
\end{figure}

We study the effects of a periodic driving. For the sake of concreteness, suppose that
a light beam comes in the $x$ direction, with the vector potential $\bA(t)=A_{0}[0,\cos(\omega t),\sin(\omega t+\phi)]$\footnote{The general incident angle is discussed in Supplemental Material.}. The choice $\phi=0$ and $\phi=\pi$ corresponds to right-handed and left-handed circularly polarized light (CPL), respectively.
The electromagnetic coupling is given by  $\mathcal{H}(\bk)
\rightarrow\mathcal{H}[\bk+e\bA(t)]$. The full Hamiltonian is time periodic; thus, it can be expanded as
$\mathcal{H}(t,\bk)=\sum_{n}\mathcal{H}_{n}(\bk)e^{in\omega t}$ with
\begin{eqnarray}
\mathcal{H}_{0}(\bk)&=&[m-Be^{2}A_{0}^{2}-Bk^{2}]\tau_{x}+vk_{z}\tau_{z},\nonumber\\
\mathcal{H}_{\pm1}(\bk)&=&-eA_{0}[2B(k_{y}\mp ie^{\pm i\phi}k_{z})\tau_{x}\pm ie^{\pm i\phi}v\tau_{z}]/2,\nonumber\\
\mathcal{H}_{\pm2}(\bk)&=&-Be^{2}A_{0}^{2}(1-e^{\pm i2\phi})\tau_{x}/4,
\end{eqnarray}
and $\mathcal{H}_{n}=0$ for $|n|>2$. In the limit where the driving frequency $\omega$ is large compared
to the other energy scales, a proper description of the system is the effective time-independent Hamiltonian\cite{Kitagawa2011,Goldman2014,Grushin2014,floquet-comment,Grushin-reply}, which reads
\begin{eqnarray}
\mathcal{H}_{\rm eff}(\bk)&=&\mathcal{H}_{0}+\sum_{n\geq1}\frac{[\mathcal{H}_{+n},\mathcal{H}_{-n}]}{n\omega}+\mathcal{O}(\frac{1}{\omega^{2}})
\nonumber\\
&=&[\tilde{m}-Bk^{2}]\tau_{x}+vk_{z}\tau_{z}+\lambda k_{y}\tau_{y}+\cdots,\quad\label{weyl}
\end{eqnarray}
where $\lambda=-2e^{2}BvA_{0}^{2}\cos\phi/\omega$ and
$\tilde{m}=m-Be^{2}A_{0}^{2}$.  The main effect of the driving
manifests in the $\lambda k_{y}\tau_{y}$ term, which is nonzero as
long as $\cos\phi\neq 0$, though circular polarizations ($\phi=0$ or
$\pi$) maximize $|\lambda|$. The energy spectra of $\mathcal{H}_{\rm
eff}$ are given by
\begin{eqnarray}
\tilde{E}_{\pm,\bk}=\pm
\sqrt{[\tilde{m}-Bk^{2}]^{2}+v^{2}k_{z}^{2}+(\lambda k_{y})^{2}}. \label{5}
\end{eqnarray}
Once $\lambda\neq0$, the energy spectra immediately become gapped except
at the two Weyl points $\bK_{\pm}=\pm(\sqrt{\tilde{m}/B},0,0)$ (see Fig.\ref{sketch}). We can expand $\mathcal{H}_{\rm eff}$
around the Weyl points as $\mathcal{H}_\pm(\bq)=\sum_{ij}v_{ij}q_{i}\tau_{j}$
with $\bq=\bk-\bK_\pm$ referring to the momentum relative to the gapless points, with $v_{xx}=\mp 2\sqrt{\tilde{m}B}, v_{yy}=\lambda, v_{zz}= v$, and all other matrix entries are $0$.
The chirality of the Weyl points at $\bK_\pm$ is
$\chi_{\pm}\equiv{\rm sgn}[\det(v_{ij})]=\pm \text{sgn}(\cos\phi)$. The appearance of $\cos\phi$ implies that the
chirality has a simple dependence on the handedness of the incident laser beam; thus, reversing the handedness causes the reversing of the chirality of Weyl points. Moreover, the locations of the Weyl points are tunable by changing the direction of the incident laser beams. For instance, the two Weyl points are located at $\pm(0,\sqrt{\tilde{m}/B},0)$ if the laser beam is along the $y$ direction.

If one considers a nodal ring of generic shape and the incident light along the $x$ direction, one can show that Weyl points are created around the local maxima and local minima of $k_x$ on the nodal ring, and the chirality of the Weyl point is opposite on the maxima and minima. Since the numbers of local minima and local maxima are equal on a ring,  the Nielsen-Ninomiya theorem\cite{NIELSEN1981} stating the equality of the numbers of Weyl points with opposite chirality is automatically satisfied.

\begin{figure}
\subfigure{\includegraphics[width=4.25cm, height=4cm]{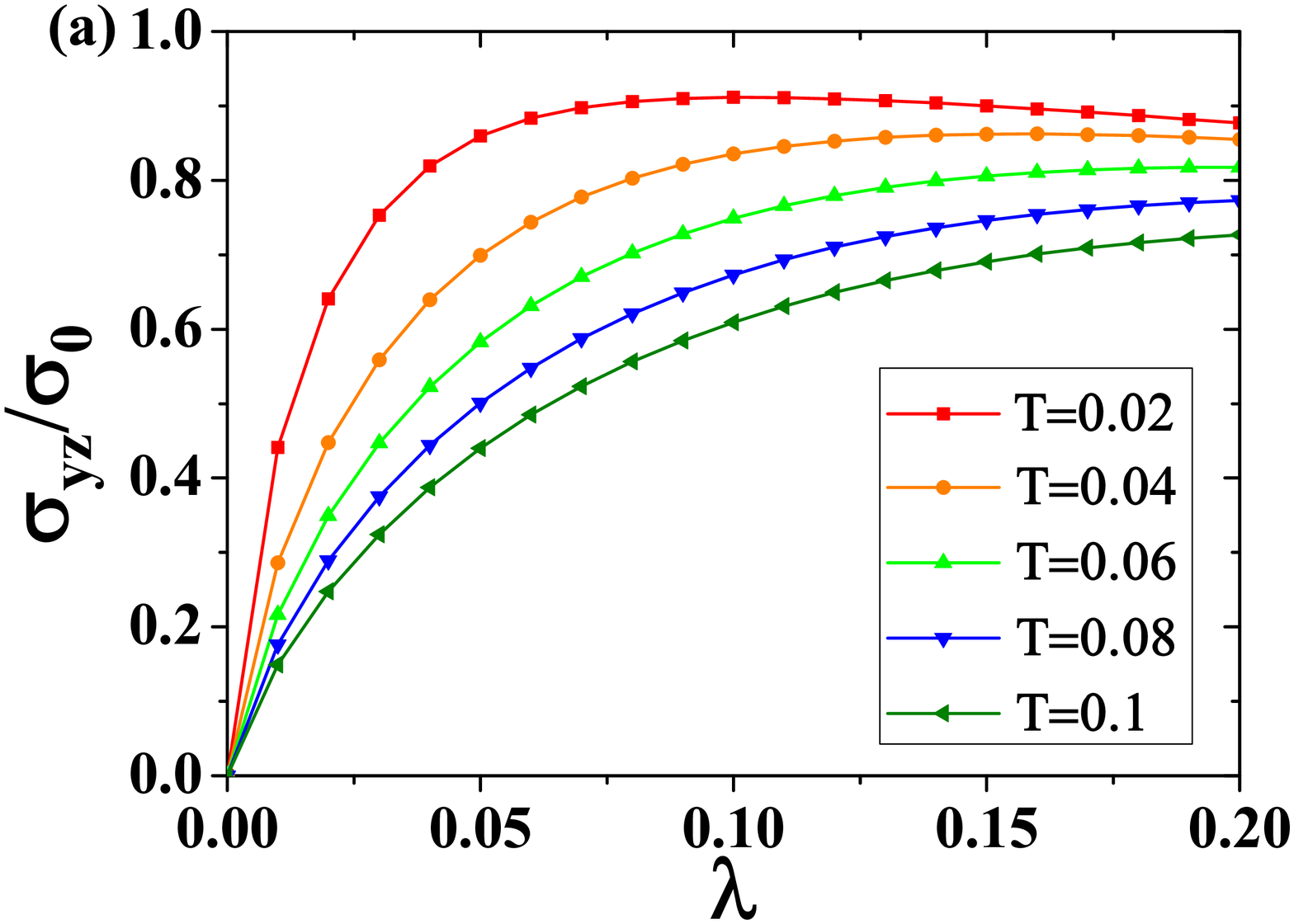}}
\subfigure{\includegraphics[width=4.25cm, height=4cm]{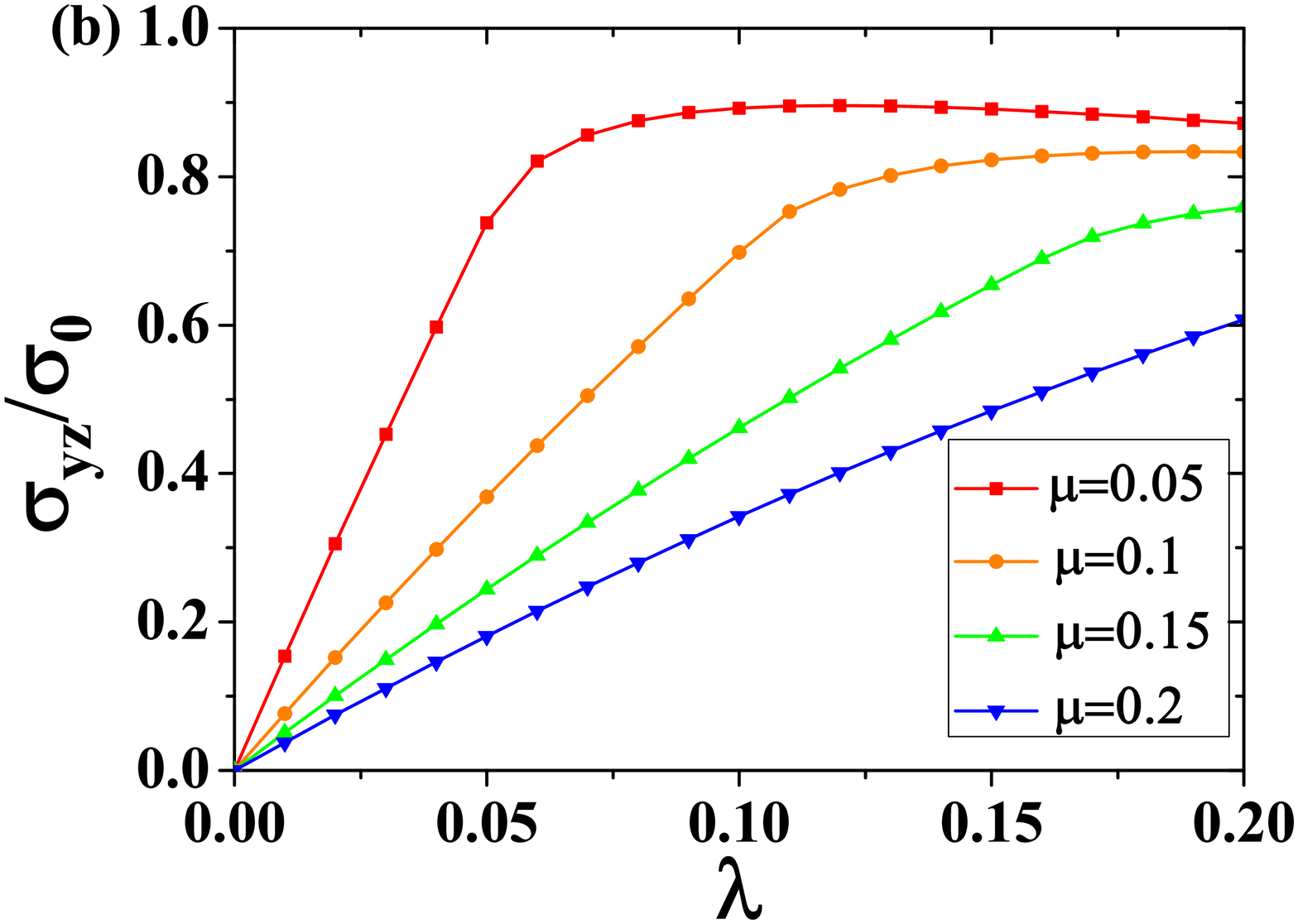}}
\caption{ The dependence of Hall conductivity on temperature ($T$),
chemical potential ($\mu$), and the intensity of the incident light beam
$\lambda$. The common parameters in use are $m=1$, $B=1$, $\omega=2$, $\phi=\pi$, $v=1$, $\epsilon_0=0$, and we have defined the shorthand notation $\sigma_{0}=\frac{e^{2}}{\pi h}\sqrt{\frac{m}{B}}$.
(a) $\sigma_{yz}$  at $\mu=0$ as a function of $\lambda$ for several values of $T$; (b) $\sigma_{yz}$  at $T=0$ as a function of $\lambda$ for several values of $\mu$. }  \label{hall}
\end{figure}

{\it Anomalous Hall effect (AHE).---}
One of the significant consequences of the topological transition from NLSM to WSM is the emergence of the AHE characterized by a nonzero
Hall conductivity.
The conductivity can be obtained from the linear response theory\cite{Oka2009}, which leads to
\begin{eqnarray}
\sigma_{\mu\nu}=e^2\epsilon_{\mu\nu\rho} \int\frac{d^3k}{(2\pi)^3}\sum_\alpha f_\alpha(k)[\nabla_{\bf k}\times\mathcal{A}_\alpha({\bf k})]_\rho
\end{eqnarray}
where $\mu,\nu,\rho=x,y,z$ and $\epsilon_{\mu\nu\rho}=\pm 1$ for the even
(odd) permutation of $(x,y,z)$,  $\alpha\equiv (i,n)$, with $i$
referring to the original band index and $n$ referring to the Floquet
index\cite{Oka2009}; $\mathcal{A}_{\alpha}(\bk)$ is the Berry
connection, and $f_\alpha$ is the occupation function. Thus the Hall
conductivity depends not only on the Berry curvature but also on the
fermion occupation, the latter of which is nonuniversal, being
dependent on details of the systems (e.g. the coupling between the
systems and the bath). Below we  focus on the case in which the
occupation is close to equilibrium, namely,
$f_{(i,n)}=\delta_{n0}n_i(E_{i,\bk})$, where $n_i(E_{i,\bk})$ is the
Fermi-Dirac distribution. For the incident light beams along the $x$
direction, the interesting component of the Hall conductivity is
$\sigma_{yz}$, which can be found as (Supplemental Material)
\begin{eqnarray}
\sigma_{yz}(T,\mu,\lambda)
&=&\frac{e^{2}}{2}\int\frac{d^{3}k}{(2\pi)^{3}}[\hat{\bd}\cdot(\frac{\partial
\hat{\bd}}{\partial k_{y}}\times\frac{\partial \hat{\bd}}{\partial k_{z}})](n_{+}-n_{-})\nonumber\\
&=&\frac{e^{2}}{2}\int\frac{d^{3}k}{(2\pi)^{3}}\frac{\lambda v(\tilde{m}-2Bk_{x}^{2}+Bk^{2})}
{\tilde{E}_{+,\bk}^{3}}(n_{+}-n_{-}), \label{AHE}
\end{eqnarray}
where $\hat{\bd}=\bd/|\bd|$ with
$\bd=\text{Tr}[\vec{\tau}\mathcal{H}_{\rm eff}]/2$, and $n_{\pm}=1/[
e^{(\tilde{E}_{\pm,\bk}-\mu)/T}+1]$ is the Fermi-Dirac distribution
with $\mu$ denoting the chemical potential.  For nonzero temperature
($T\neq0$) or doped system ($\mu\neq0$), analytic simplification of
Eq.(\ref{AHE}) is not available, and we need to treat Eq.(\ref{AHE})
numerically(see Fig.\ref{hall}). From Fig.\ref{hall}(a) we can see that
$\sigma_{yz}(T,0,\lambda)$ increases and saturates as $\lambda$ is
increased. The saturation occurs at smaller $\lambda$ when $T$ is
lower. As $T\rightarrow 0$, the $\sigma_{yz}$-$\lambda$ curve
approaches a step function, jumping from $\sigma_{yz}=0$ at
$\lambda=0$ to a nonzero value $\sigma_{yz}(0,0,\lambda\neq 0)$ at
$\lambda>0$. Taking $\mu=T=0$ in Eq.(\ref{AHE}) leads to
\begin{eqnarray}
\sigma_{yz}(0,0,\lambda\neq 0) =\frac{e^{2}}{\pi
h}\sqrt{\frac{m}{B}-e^{2}A_{0}^{2}}\text{sgn}(\cos\phi),\label{conductivity}
\end{eqnarray}
where we have restored the Planck constant $h$. It is readily seen
that $\sigma_{yz}(0,0,\lambda\neq 0)$ is proportional to the distance
between the two Weyl points. The Hall conductivity can be easily
tuned by the incident angle of light. For instance, the nonzero
component is $\sigma_{xz}$ instead of $\sigma_{yz}$ if the
incident light is along the $y$ direction.

This behavior is remarkably different from the light-induced Hall
effect in Weyl semimetals\cite{Chan2016hall}. The proposal of
Ref.\cite{Chan2016hall} is to separate existing Weyl points in WSMs
by light, while ours is to create Weyl points from NLSMs. The former
is a second-order effect proportional to $A_0^2$, while the AHE in
the present work is a zeroth-order effect, which should be much more
pronounced in experiment. In general, the distance between the two
photoinduced Weyl points is of the order of $2\pi/a$,  $a$ referring
to the lattice constant; thus, we have the estimation $\sigma_{yz}\sim
(e^2/h)(2\pi/a)$.

{\it Fermi arc as the descendent of drumhead states.---} The
dispersion of the surface states of the NLSM takes the shape of a
drumhead or a bowl, i.e., a nearly flat band bounded by the
projection of the bulk nodal line to the surface Brillouin zone. When
$\epsilon_0= 0$, the dispersion becomes exactly flat. Since the NLSM
is driven to a WSM phase in our study, it is a natural question how a
Fermi arc comes from a drumhead (or bowl). We consider a
semi-infinite geometry, namely, the sample occupies the entire $z>0$
half-space. The ${\bf k}$-space energy eigenvalue problem is
translated to the real space as $\mathcal{H}_{\rm
eff}(k_{x},k_{y},-i\partial_{z})\Psi(x,y,z)=E(k_{x},k_{y})\Psi(x,y,z)$, under
the boundary conditions $\Psi(z=0)=0$ and $\Psi(z\rightarrow+\infty)=0$.

\begin{figure}
\subfigure{\includegraphics[width=7cm, height=5cm]{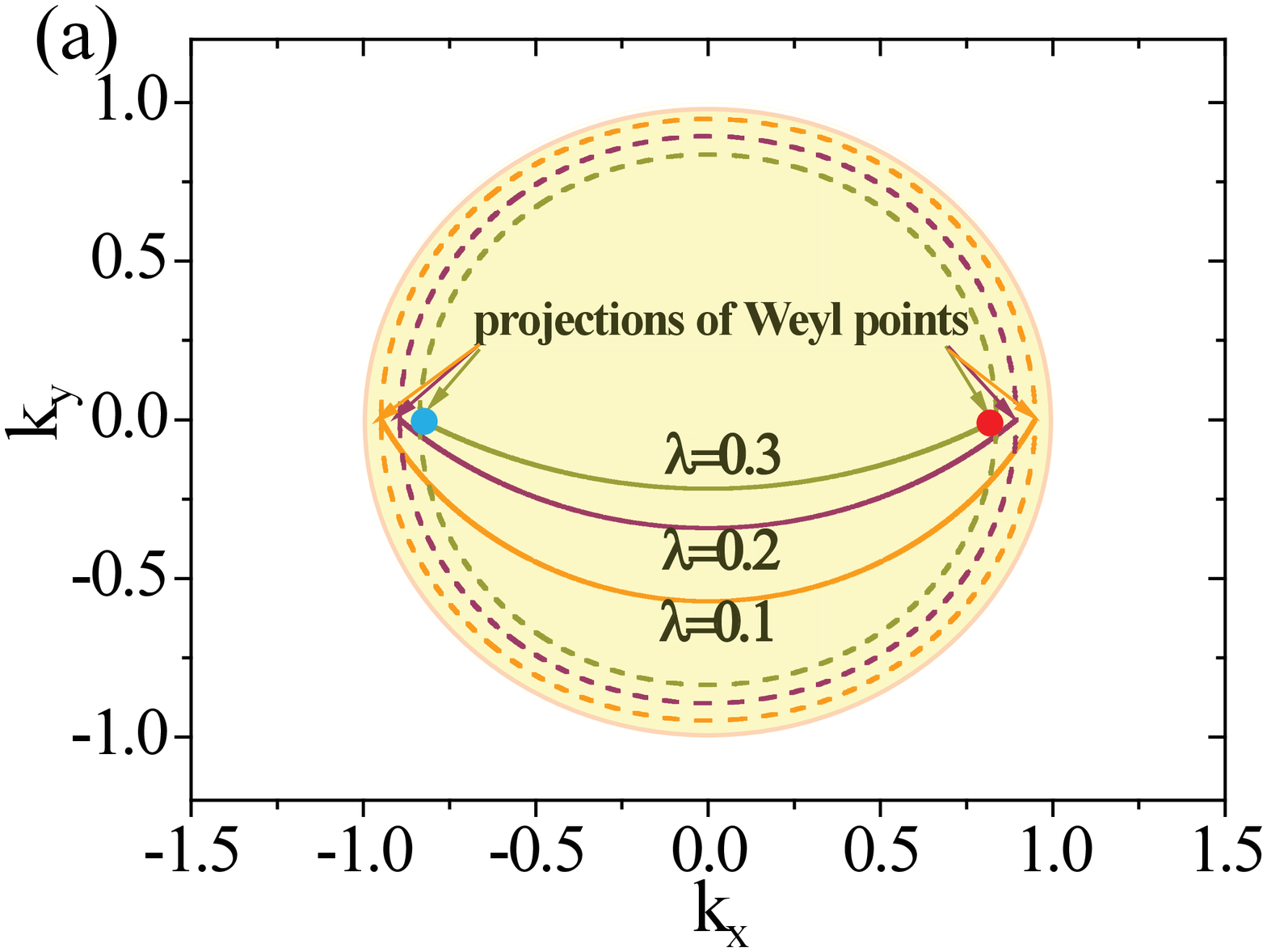}}
\subfigure{\includegraphics[width=7cm, height=5cm]{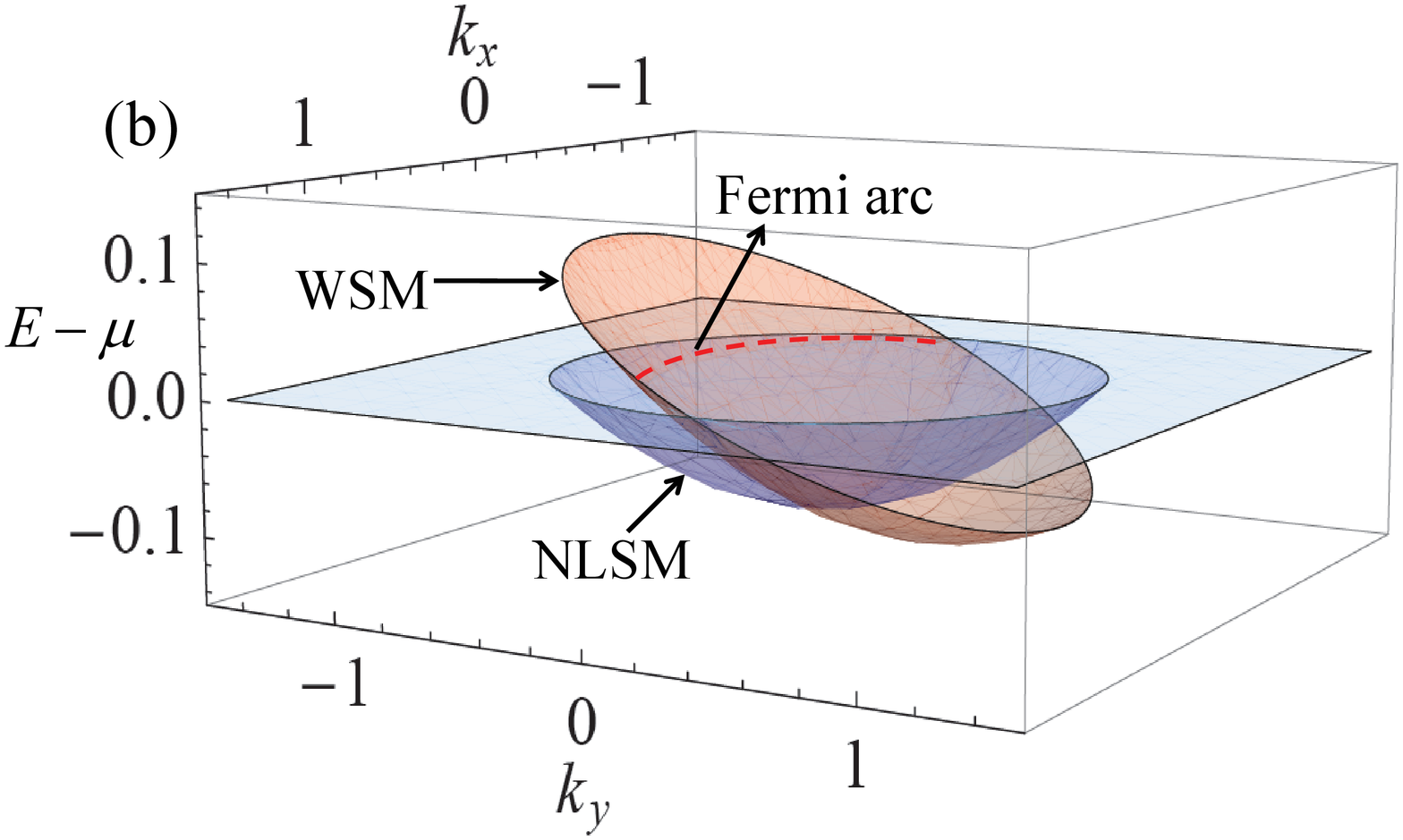}} \caption{
The surface states of the NLSM and the Floquet WSM. The values of
parameters are taken to be $m=B=1$, $\omega=2$, $\phi=\pi$, $v=1$,
and $\epsilon_0(k)=0.1(k_x^2+k_y^2)$. The chemical potential $\mu$
has been tuned to the Weyl band-touching points. (a)  The shadow area
($k_x^2+k_y^2<m/B$) represents the drumhead surface states of the
pristine NLSM, and the dashed lines enclose the area of surface
states of the Floquet WSM($k_x^2+k_y^2<\tilde{m}/B$), with the three
colors referring to three values of $\lambda$, as indicated in the
figure. The solid curves are the Fermi arcs connecting the
projections of the two Weyl points onto the surface Brillouin zone.
(b) The surface state dispersions of both the pristine NLSM and the
Floquet WSM (with the driving parameter $\lambda=0.1$). Tilting the
surface state dispersion of the NLSM leads to that of the WSM. The
intersection of the dispersion of WSM and the $E=\mu$ plane is the
Fermi arc. } \label{arc}
\end{figure}

We find that the surface mode wave function takes the form of \bea
\Psi(x,y,z) = \mathcal{N}e^{ik_xx}e^{ik_yy}\sin(\kappa z)e^{-\gamma
z}\chi, \eea with \bea E(k_x,k_y) = -\lambda k_y
+\epsilon_0(k_x,k_y), \label{dispersion} \eea where
$\chi=(1,-i)^T/\sqrt{2}$, $\mathcal{N}$ is a normalization factor,
and $\gamma =\frac{v}{2B}, \quad \kappa = \frac{1}{2B}\sqrt{
4B[\tilde{m}  - B(k_x^2+k_y^2)] -v^2}$. Implicit here is that
$4B[\tilde{m}  - B(k_x^2+k_y^2)] -v^2>0$, so that $\kappa$ is
real valued; in the cases of $4B[\tilde{m}  - B(k_x^2+k_y^2)]
-v^2<0$, the solution can be obtained by replacing $\sin(\kappa z)$
by $\sinh(|\kappa|z)$, namely, $\Psi(x,y,z)
=\mathcal{N}e^{ik_xx}e^{ik_yy}\sinh(|\kappa| z)e^{-\gamma z}\chi$. It
is found that the normalizability requires \bea
k_x^2+k_y^2<\tilde{m}/B, \label{region} \eea which determines the
region where the surface modes exist. It is notable that the
dispersion given by Eq.(\ref{dispersion}) shows a chiral nature in
the $y$ direction. As a comparison, we note that the surface states
of the NLSM can be recovered by letting $\lambda=0$, which, according
to Eq.(\ref{dispersion}), leads to the dispersion of the
drumhead or bowl states: $E(k_x,k_y)=\epsilon_0(k_x,k_y)$. The region of
the drumhead or bowl is given by Eq.(\ref{region}) with $\tilde{m}$
replaced by $m$, namely, $k_x^2+k_y^2<m/B$. An illustration is given
in Fig.\ref{arc}(a), in which the surface state of the pristine NLSM
is represented by the shadow area, and the surface states of the
Floquet WSM for three values of intensity of incident laser are
enclosed by the three dashed lines with different colors. Because of the
nonzero $\lambda$ generated by the incident laser, which  enters
Eq.(\ref{dispersion}), the dispersion of the drumhead or bowl states of
the pristine NLSM becomes tilted (see Fig.\ref{arc}(b)).

{\it Spinful NLSMs.---} SOC can change the size
and location of nodal lines or gap them out. Let us first consider
the first case. For concreteness, we consider a model Hamiltonian
$\mathcal{H}_{\rm s} (\bk)= (m-Bk^2)\tau_x + vk_z\tau_z +\Delta_{\rm
so}\tau_x s_z$, where $s_z$ is the Pauli matrix for spin, and
$\Delta_{\rm so}$ quantifies SOC. This model hosts two nodal lines,
and is relevant to the NLSM candidate
${\mathrm{TlTaSe}}_{2}$\cite{Bian2015TlTaSe}. Using the method of the
previous sections, we can find the effective Floquet Hamiltonian
$\mathcal{H}_{s,\rm eff} (\bk) = (m-Bk^2)\tau_x + vk_z\tau_z
+\Delta_{\rm so}\tau_x s_z +\lambda k_y\tau_y$, with spectrum $E=\pm
\sqrt{(\tilde{m}\pm\Delta_{\rm so} - Bk^2)^2+v^2 k_z^2 +\lambda^2
k_y^2}$, in which the expressions for $\lambda$ and $\tilde{m}$ are
the same as given below Eq.(\ref{weyl}). Thus the system is driven to
the WSM phase with four Weyl points at ${\bf K}^\pm_1 =\pm
(\sqrt{(\tilde{m}+\Delta_{\rm so})/B}, 0, 0)$ and ${\bf K}^\pm_2 =\pm
(\sqrt{(\tilde{m}-\Delta_{\rm so})/B}, 0, 0)$.

{\it Effect of a small gap.---} In several candidates of NLSMs, the
nodal line can be gapped out if a small spin-orbit
coupling\cite{Yu2015,Kim2015,Lu2016Node} is included. Below we
show that such a small gap introduces a threshold laser intensity.
Let us consider the model \bea \mathcal{H}_{\rm g} (\bk)=
(m-Bk^2)\tau_x + vk_z\tau_z +\lambda_{\rm so}\tau_y s_z, \eea where
$s_z$ refers to the $z$ component of spin, and the $\lambda_{\rm
so}\tau_y s_z$ term induces an energy gap $2|\lambda_{\rm so}|$. With
an incident laser, the effective Hamiltonian reads \bea
\mathcal{H}_{g,\rm eff} (\bk) = (m-Bk^2)\tau_x + vk_z\tau_z
+\lambda_{\rm so}\tau_y s_z +\lambda k_y\tau_y, \eea with spectrum
$E=\pm \sqrt{(\tilde{m}  - Bk^2)^2+v^2 k_z^2 + (\lambda k_y\pm
\lambda_{\rm so})^2}$. There are clearly four generated Weyl points,
located at ${\bf Q}^\pm_1 =(\pm\sqrt{\tilde{m}/B-(\lambda_{\rm
so}/\lambda)^2}, \lambda_{\rm so}/\lambda, 0)$, and ${\bf Q}^\pm_2
=(\pm\sqrt{\tilde{m}/B-(\lambda_{\rm so}/\lambda)^2}, -\lambda_{\rm
so}/\lambda, 0)$. It is readily seen that the main effect of the gap
 is a threshold laser intensity $\lambda_{\rm th}=\sqrt{\lambda_{\rm
so}^2B/\tilde{m}}$, above which the Weyl points can be generated.
Below the threshold laser intensity the system is gapped.

{\it Experimental considerations.---} Among other possibilities, it
should be feasible to test our predictions in the film of NLSMs. The
incident light has a finite penetration depth $\delta$, and it is
 most efficient to measure the Hall conductivity in film with
thickness $\sim\delta$. The estimation of $\delta$ becomes simple
when the Fermi velocities along the $z$ direction and the $x-y$
direction are the same, namely, when $2\sqrt{mB}=v$, and the optical
absorption rate can be straightforwardly estimated following
Ref.\cite{nair2008fine}, leading to $\delta\sim
\sqrt{B/(m\alpha^2)}$, where $\alpha\sim 1/137$ is the fine structure
constant. Since $\sqrt{B/m}$ is the inverse of the radius of the
nodal ring, we expect that it is of the order of a few lattice
constants, thus $\delta\sim$ several hundred lattice constants.
Taking the NLSM candidate ZrSiS as an example, we have estimated that
a film of thickness $100$nm and size $100 \mu{\rm m} \times 100 \mu{\rm m}$
can generate a Hall voltage$\sim 1$ mV with a dc current of $100$ mA
(Supplemental Material), which can be readily measured experimentally.

In the pump-probe experiments\cite{wang2013observation}, $2e^2 v^2 A_0^2/\omega\sim 50$meV is attainable at photon energy $\hbar\omega=120$meV (Ref.\cite{wang2013observation}); therefore, if we take $v^2/B\sim 0.5$eV, we have $|\lambda|\sim 0.1v$. Since we have shown that the Fermi velocity
of the Fermi arc is $\lambda$, we can see that, within the current experimental feasibility, it can reach a tenth of the bulk Fermi velocity of the
NLSMs. This can be detected by the pump-probe angle-resolved photoemission spectroscopy (ARPES)\cite{wang2013observation,mahmood2016selective}.

{\it Conclusions.---} We have shown that NLSMs are tailor-made materials for optically
creating Floquet Weyl points. Remarkably, polarized light with infinitesimal intensity
is sufficient in principle. The resultant
Floquet WSMs have a large AHE controllable by the laser beams, e.g.,
reversing the handedness of the incident light changes the sign of
the anomalous Hall conductivity.  The Fermi arcs of the Floquet WSMs
have an appealing interpretation, namely, they come from tilting the
drumhead surface dispersion of NLSMs. The photoinduced Fermi arcs and
bulk Weyl points can be detected in pump-probe ARPES, and the AHE
can be measured in transport experiments. Our proposal can be generalized to cold-atom systems\cite{Xu2016Dirac} by
shaking lattice\cite{jotzu2014experimental,parker2013direct,Zheng2014}.

Apart from potential applications of the large and tunable AHE in high-speed electronics, the tunability of the Floquet WSMs
also facilitates the future investigations of many novel physics therein by techniques absent in the static systems (e.g. spatial modulation of light \cite{Katan2013modulated}). From a broader perspective, our work suggests that Floquet topological semimetals are fruitful platforms in the study of topological matters.

\emph{Acknowledgements.} This work is supported by NSFC under Grant
No. 11304175.

Note added: After initial submission of this letter, there appeared two related
preprints\cite{Chan2016type,Narayan2016}. In the overlapping parts, their conclusions
are consistent with ours.

\bibliography{dirac}

\vspace{8mm}

{\bf Supplemental Material}

\vspace{4mm}

This supplemental material contains: (i) The creation of Weyl points
in nodal line semimetals by circularly polarized light with a general
incident angle, (ii) The derivation of the anomalous Hall
conductivity, and (iii) Details of experimental estimations.

\section{Creation of Floquet Weyl points by incident light in an arbitrary direction }

In the main text we have considered the case that the incident light is along the $x$ direction, which is in the plane of nodal line. In this supplemental material we treat the cases of general incident angles. We shall see that the main results are qualitatively the same, except when the light is along the $z$ direction, for which Floquet Weyl points cannot be created.

\begin{figure}
\includegraphics[width=6cm, height=4cm]{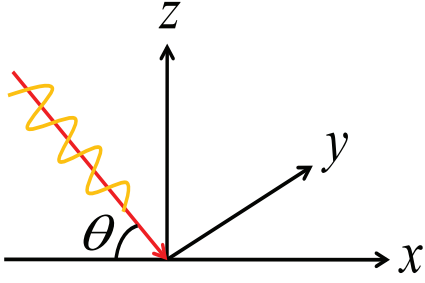}
\caption{ Sketch of the incident angle.}  \label{incident}
\end{figure}

As the Hamiltonian has rotation symmetry in $x$-$y$ plane, we can always rotate along the $z$ axis so that the incident light is parallel to the $x-z$ plane. Let us denote the angle between the incident light and the $x$ axis as $\theta$ (see Fig.\ref{incident}).
The vector potential is
\begin{eqnarray}
\bA(t)=A_{0}(\,\sin\theta\sin(\omega t+\phi),\,\cos(\omega t),\,\cos\theta\sin(\omega t+\phi)\,),
\end{eqnarray} where $\phi=0$ and $\phi=\pi$ corresponds to right-handed and left-handed circularly polarized light, respectively.
The time-periodic Hamiltonian can be expanded as  $\mathcal{H}(\bk,t)=\sum_{n}\mathcal{H}_{n}(\bk)e^{in\omega t}$, with the Fourier components
\begin{eqnarray}
\mathcal{H}_{0}(\bk)&=&[m-Be^{2}A_{0}^{2}-Bk^{2}]\tau_{x}+vk_{z}\tau_{z},\nonumber\\
\mathcal{H}_{\pm1}(\bk)&=&-eA_{0}[B(  k_{y} \mp ie^{\pm i\phi}k_{x}\sin\theta\mp ie^{\pm i\phi}k_{z}\cos\theta)\tau_{x}\nonumber\\
&&\pm(ie^{\pm i\phi}v \cos\theta/2)\tau_{z}],\nonumber\\
\mathcal{H}_{\pm2}(\bk)&=&-Be^{2}A_{0}^{2}(1-e^{\pm i2\phi})\tau_{x}/4.
\end{eqnarray}
and $\mathcal{H}_{n}=0$ for $|n|>2$. The effective time-independent Hamiltonian can be obtained as
\begin{eqnarray}
\mathcal{H}_{\rm eff}(\bk)&=&\mathcal{H}_{0}+\sum_{n\geq1}\frac{[\mathcal{H}_{+n},\mathcal{H}_{-n}]}{n\omega}+\mathcal{O}(\frac{1}{\omega^{2}})
\nonumber\\
&=&[m-Be^{2}A_{0}^{2}-Bk^{2}]\tau_{x}+vk_{z}\tau_{z}+\lambda\cos\theta k_{y}\tau_{y},
\end{eqnarray}
where $\lambda=-2Bve^{2}A_{0}^{2}\cos\phi/\omega$. Therefore, tuning the incident direction from $\theta=0$ (i.e. the $x$ direction) to a general $\theta$ changes the $y$-th Fermi velocity of the Floquet Weyl points to $\lambda\cos\theta$.  When $\cos\theta\neq0$, i.e.,
the incident light is not in $z$ direction, two Weyl points are created at
$\bK_{\pm}=(\pm\sqrt{\frac{m}{B}-e^{2}A_{0}^{2}},0,0)$.
Therefore, as long as the incident light is not perpendicular to the plane of nodal
line/ring, the nodal line semimetal will be driven to Weyl semimetal by a circularly polarized light.

\section{Deviation of the formulas of anomalous Hall conductivity}

According to the Floquet theory(see e.g. Ref.\cite{Kitagawa2011}),
the wavefunction $\Psi_\alpha({\bf k}, t)$ that satisfies the
time-dependent Schrodinger equation $i\partial_t \Psi_\alpha({\bf k},
t) = \mathcal{H}(\bk,t) \Psi_\alpha({\bf k}, t)$ can be written as
$\Psi_\alpha({\bf k}, t)= e^{-i\epsilon_\alpha t}
\Phi_{\alpha}(\bk,t)$, in which $\epsilon_\alpha$ is the so-called
quasi-energy and $\Phi_{\alpha}(\bk,t)$ is periodic, namely,
$\Phi_{\alpha}(\bk,t+T)=\Phi_{\alpha}(\bk,t)$, $T=2\pi/\omega$ being
the driven period. The periodic function $\Phi_{\alpha}(\bk,t)$
satisfies $[\mathcal{H}(\bk,t) -i\partial_t] \Phi_\alpha({\bf k}, t)
=
 \epsilon_\alpha\Phi_\alpha({\bf k}, t)$.

Since $\Phi_{\alpha}(\bk,t)$ is periodic, it can be expanded as
$\Phi_{\alpha}(\bk,t)=\sum_{n}e^{in\omega
t}\Phi_{\alpha}^{(n)}(\bk)$, in which the Fourier components
$\Phi_{\alpha}^{(n)}(\bk)$ satisfy the Floquet equation
\begin{eqnarray}
[\varepsilon_{\alpha}(k)-n\omega]\Phi_{\alpha}^{(n)}(\bk)=\sum_{m}\mathcal{H}_{n-m}(\bk)\Phi_{\alpha}^{(m)}(\bk)\label{S2}
\end{eqnarray}
with $\mathcal{H}_{n-m}(\bk)=\frac{1}{T}\int_{0}^{T}dte^{i(m-n)\omega
t}\mathcal{H}(\bk,t)$. In the off-resonant regime ($\omega$ much
larger than other energy scale), based on perturbation
theory\cite{Kitagawa2011}, it is found that
\begin{eqnarray}
&&\mathcal{H}_{\rm eff}(\bk)\Phi_{\alpha}^{(0)}(\bk)=E_{\alpha}(\bk)\Phi_{\alpha}^{(0)}(\bk),\nonumber\\
&&\Phi_{\alpha}^{(n)}(\bk)=-\frac{\mathcal{H}_{n}}{n\omega}\Phi_{\alpha}^{(0)}(\bk)\quad \text{for}\, n\neq0.\label{S4}
\end{eqnarray}
where
\begin{eqnarray}
\mathcal{H}_{\rm eff}({\bf k})= \mathcal{H}_{0}+\sum_{n\geq1}\frac{[\mathcal{H}_{+n},\mathcal{H}_{-n}]}{n\omega}+\mathcal{O}(\frac{1}{\omega^{2}})
\end{eqnarray}

It follows from the linear response theory\cite{Oka2009} that the
formula for the Hall conductivity is given by
\begin{eqnarray}
\sigma_{\mu\nu}=e^{2}\epsilon_{\mu\nu\rho}\int\frac{d^{3}k}{(2\pi)^{3}}
\sum_{\alpha}f_{\alpha}(k)[\nabla_{\bk}\times\mathcal{A}_{\alpha}(\bk)]_{\rho},\label{S1}
\end{eqnarray}
where $\mathcal{A}_{\alpha}(\bk)=-i \ll
\Phi_{\alpha}(\bk)|\nabla_{\bk}|\Phi_{\alpha}(\bk)\gg\equiv\frac{1}{T}\int_{0}^{T}dt
\langle\Phi_{\alpha}(\bk,t)|\nabla_{\bk}|\Phi_{\alpha}(\bk,t)\rangle$.
The Berry connection $\mathcal{A}_{\alpha}(\bk)$ can also be written
as
\begin{eqnarray}
\mathcal{A}_{\alpha}(\bk)=\sum_{n}\mathcal{A}_{\alpha,n}(\bk)=
-i\sum_{n}\langle\Phi_{\alpha}^{(n)}(\bk)|\nabla_{\bk}|\Phi_{\alpha}^{(n)}(\bk)\rangle\label{S3}
\end{eqnarray}
As has been obtained in the main text, $\mathcal{H}_{\rm eff}(\bk)$
takes the following form,
\begin{eqnarray}
\mathcal{H}_{\rm eff}(\bk)=[\tilde{m}-Bk^{2}]\tau_{x}+vk_{z}\tau_{z}+\lambda k_{y}\tau_{y}. \label{S5}
\end{eqnarray}
It is readily found that $E_{\alpha=\pm}(\bk)=\pm\sqrt{[\tilde{m}-Bk^{2}]^{2}+v^{2}k_{z}^{2}+\lambda^{2} k_{y}^{2}}$,
and $\Phi_{\alpha}^{(0)}(\bk)$ are given by
\begin{eqnarray}
\Phi_{+}^{(0)}(\bk)=\left(\begin{array}{c}
                      \cos\frac{\theta_{\bk}}{2} \\
                      \sin\frac{\theta_{\bk}}{2}e^{i\varphi_{\bk}}
                    \end{array}\right),
\Phi_{-}^{(0)}(\bk)=\left(\begin{array}{c}
                      \sin\frac{\theta_{\bk}}{2}e^{-i\varphi_{\bk}} \\
                      -\cos\frac{\theta_{\bk}}{2}
                    \end{array}\right),        \label{S6}
\end{eqnarray}
where $\theta_{\bk}=\arccos\frac{vk_{z}}{E_{+}(\bk)}$ and $\varphi_{\bk}=
\arctan\frac{\lambda k_{y}}{\tilde{m}-Bk^{2}}$. For the
case that the occupation is close to equilibrium, namely,
$f_{\pm,n}(k)=\delta_{n,0}n_{\pm}(E_{\pm}(\bk))$, where $n_{\pm}(E_{\pm}(\bk))=1/(e^{(E_{\pm}(\bk)-\mu)/T}+1)$
is the Fermi-Dirac distribution. With the $\mathcal{O}(1/\omega^2)$ terms omitted, the Hall conductivity takes the following form:
\begin{eqnarray}
\sigma_{\mu\nu}=e^{2}\epsilon_{\mu\nu\rho}\int\frac{d^{3}k}{(2\pi)^{3}}\sum_{i=\pm}n_{i}(E_{i}(\bk))
[\nabla_{\bk}\times\mathcal{A}_{i,0}(\bk)]_{\rho}.\label{S7}
\end{eqnarray}
A combination of Eq.(\ref{S3}) and  Eq.(\ref{S6}) gives
\begin{eqnarray}
\mathcal{A}_{+,0}(\bk)=-\mathcal{A}_{-,0}(\bk)=\sin^{2}\frac{\theta_{\bk}}{2}\nabla_{\bk}\varphi_{\bk},
\end{eqnarray}
thus Eq.(\ref{S7}) can be rewritten as
\begin{eqnarray}
\sigma_{\mu\nu}=\frac{e^{2}}{2}\epsilon_{\mu\nu\rho}\int\frac{d^{3}k}{(2\pi)^{3}}\sin\theta_{\bk}
[\nabla_{\bk}\theta_{\bk}\times\nabla_{\bk}\varphi_{\bk}]_{\rho}(n_{+}-n_{-}),
\end{eqnarray}
this formula can be further rewritten as
\begin{eqnarray}
\sigma_{\mu\nu}=\frac{e^{2}}{2}\int\frac{d^{3}k}{(2\pi)^{3}}[\hat{\bd}\cdot
(\partial_{k_{\mu}}\hat{\bd}\times\partial_{k_{\nu}}\hat{\bd})](n_{+}-n_{-}),
\end{eqnarray}
where $\hat{\bd}=(\sin\theta_{k}\cos\varphi_{\bk},\sin\theta_{k}\sin\varphi_{\bk},\cos\theta_{k})$.
The unit vector $\hat{\bd}$ can also be expressed in the form $\hat{\bd}\equiv\bd/|\bd|$ with
$\bd=\text{Tr}[\vec{\tau}\mathcal{H}_{\rm eff}]/2$. A direct calculation
leads to the final expression
\begin{eqnarray}
\sigma_{yz}=\frac{e^{2}}{2}\int\frac{dk^{3}}{(2\pi)^{3}}\frac{\lambda v(\tilde{m}-2Bk_{x}^{2}+Bk^{2})}
{[E_{+}(k)]^{3}}(n_{+}-n_{-}).
\end{eqnarray}

\section{Details of experimental estimations}

\begin{figure}
\includegraphics[width=6cm, height=4cm]{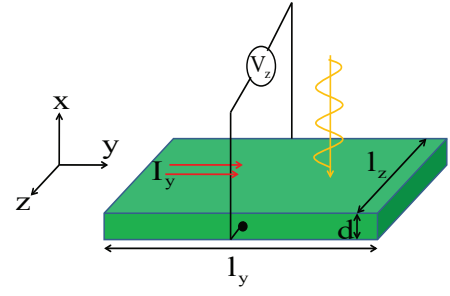}
\caption{ Sketch of the experimental setup for the measurement of
anomalous Hall effect. The nodal line lies in the $x$-$y$ plane, and
the circularly polarized light is incident along the $x$ direction.
$I_{y}$ denotes an electric current in the $y$ direction, $l_{y}$ is
the spacing between the current contacts,  $l_{z}$ is the spacing
between the voltage contacts, $d$ is the thickness of the sample, and
$V_{z}$ is the Hall voltage to be measured.}  \label{Hall}
\end{figure}

In this section, we provide more details for the estimated transport
measurements discussed in the main text. We still assume that the
nodal line lies in the $x$-$y$ plane, and the circularly polarized
light is incident along the $x$ direction. A sketch of the
experimental setup for measurement of the anomalous Hall effect is
given in Fig.\ref{Hall}. In this setup, $I_{y}$ is a constant current
in the $y$ direction, $l_{y}$ is the spacing between current contacts
and $l_{z}$ is the spacing between voltage contacts, $d$ is the
thickness of the sample, and $V_{z}$ refers to the Hall voltage to be
measured. Assuming a uniform distribution of the current through the
entire thickness,  we obtain the Hall voltage $V_{z}$ as follows
\cite{Chan2016hall}:
\begin{eqnarray}
V_{z}\approx\frac{\sigma_{yz}\delta/d}{\sigma_{yy}^{2}+(\sigma_{yz}\delta/d)^{2}}\times \frac{l_{z}}{l_{y}d}\times I_{y},\label{Ohm}
\end{eqnarray}
where $\delta$ is the penetration depth of light, given by
$\delta(\omega)=\frac{n(\omega)\epsilon_{0}c}{\text{Re}\sigma_{xx}(\omega)}$,
where $n(\omega)$ is the refraction index of the sample,
$\epsilon_{0}$ the permittivity of vacuum, $c$ the speed of light,
and $\text{Re}\sigma_{xx}(\omega)$ denotes the real part of the
optical conductivity in the $x$ direction. For nodal line semimetal,
if we assume that the Fermi velocities in the two transverse
directions are the same, we have
$\text{Re}\sigma_{xx}(\omega)=\frac{e^{2}}{h}\frac{\pi}{8}\sqrt{\frac{m}{B}}$.

In the following, we take the experimentally-confirmed nodal line
semimetal ZrSiS as a concrete example to estimate the Hall
conductivity $\sigma_{yz}$ and the Hall voltage $V_{z}$. For a
circularly polarized light parallel to the $x$-$y$ plane ( or the
$ab$ plane) of ZrSiS, the distance between induced Weyl points is
$\Delta\approx2\pi/a$ (i.e. the size of the nodal line in ZrSiS is
very large) with $a\simeq3.5${\AA} the lattice constant in $x$-$y$
plane. Thus
\begin{eqnarray}
\sigma_{yz}=\frac{e^{2}}{h}\frac{\Delta}{2\pi}\approx
1.0\times10^{5}\Omega^{-1}m^{-1}.\label{HC}
\end{eqnarray}
To be close to the experimental condition\cite{wang2013observation},
we take the frequency of light to be $\hbar\omega=120$ meV. Let us
take the refraction index to be $n(\omega)\approx 3$, then the
penetration depth is\cite{Chan2016hall}
\begin{eqnarray}
\delta&\simeq&\frac{n(\omega)\epsilon_{0}c}{(\pi e^{2}/16h)\Delta}\nonumber\\
&\approx&\frac{3\times8.85\times10^{-12}\times3\times10^{8}}{(\frac{1.6^{2}\times10^{-38}}{6.63\times10^{-34}})\times(3.14^{2}/8)}a\nonumber\\
&\approx&167a\approx586\text{{\AA}}.\label{PD}
\end{eqnarray}
We shall use the measured value\cite{Singha2016} of the dc
conductivity $\sigma_{xx}$ at 300 K, which is
$\sigma_{xx}\approx6.6\times10^{6}$ $\Omega^{-1}m^{-1}$. Let us take
the sample size to be $l_{y}=l_{z}=100$ $\mu$m, $d=100$ nm. Suppose
that the electric current is $I_{y}=100$ mA, then a combination of
Eq.(\ref{Ohm}), Eq.(\ref{HC}) and Eq.(\ref{PD}) yields the resultant
Hall voltage $V_{z}\approx1.3$ mV, which is easily observable in
realistic experiments.

\end{document}